\documentclass[12pt]{article}

\begin{document}

\title{Instanton constraints\\ in supersymmetric gauge theories
II.\\ $N=2$ Yang-Mills theory}
\author{N.K. Nielsen \footnote{Electronic address:
nkn@fysik.sdu.dk}\hspace{1 cm}   
\\ Physics Department, University of Southern
Denmark,\\ Odense, Denmark.}
\date{\today}
\maketitle

\begin{abstract}
The analysis in previous publications of the instanton
constraints required to produce a finite action of the
theory is carried out also for $N=2$ supersymmetric Yang-Mills
theory.\\ PACS numbers 11.15.-q, 12.60.Jv
\end{abstract}

\vspace{1 cm}

\section{Introduction} \label{Int}
Instantons play a prominent role in nonperturbative studies of
supersymme\-tric gauge theories. In the case of $N=2$
supersymmetric Yang-Mills theory, the connection from
instantons to the celebrated Seiberg-Witten solution
\cite{SeibergWitten} was established  by Finnell and Pouliot
\cite{FinnellPouliot}.  The instantons in question are constrained instantons
\cite{Affleck} since the scalar field of the theory has a nonvanishing vacuum
expectation value. The choice of constraint was shown in
\cite{NN}   to be  restricted since not all
constraints lead to a finite action instanton solution. 

In the present paper the investigation reported
in \cite{NN}  on permissible instanton constraints  is
extended to the case of $N=2$ supersymmetric
Yang-Mills theory. Here the scalar field is in the
adjoint representation. As a consequence, two different
prepotentials, obeying a  complicated set of coupled 
differential equations, are necessary for the gauge field.
Remarkably it is found, however, that the constraint can
be chosen in a gauge covariant way, and the main feature of
the result of
\cite{NN} persists:  Constraint terms in the gauge field
equations are only necessary at second and fourth order of
the scalar field vacuum expectation value. This  result actually applies to all $SU(2)$ Yang-Mills
field theories where the gauge field couples to a scalar
field in the adjoint representation, and thus also to $N=4$
supersymmetric Yang-Mills theory. 

The fermion zero modes are also investigated, and it
is found that the supersymmetric zero modes automatically
have acceptable asymptotic behaviour to all orders in the
scalar field vacuum expectation value, whereas the
superconformal zero modes (the eigenvalue of which is 
lifted) have unpermitted large-distance behaviour at first
order. 

This paper parallels one on supersymmetric QCD \cite{GHN}, and many features of
the analysis reported there persist in the present case.
Especially the analysis of the fermion zero modes turns out
to be very similar to, though simpler than, the analysis of
\cite{GHN}; for the sake of completeness it is nevertheless
presented in some detail. The analysis of the instanton
mass corrections, on the other hand, is considerably more
complicated here, though the final result 
on
constraints is the same as in \cite{NN},
\cite{GHN}.

In sec.II the action of supersymmetric $N=2$ Yang-Mills
theory and its continuation to Euclidean space are
presented, while the analysis of the instanton mass
corrections is given in sec.III and of the fermion zero
modes in sec.IV. The results are summarized in a brief
conclusion, while an appendix contains supplementary
material for sec.III.

 \section{The action}

The Lagrangian of $N=2$ supersymmetric Yang-Mills theory with gauge group $SU(2)$ is:
\begin{eqnarray}&& 
L_{N=2}=-(D^\mu  A^a)^\dagger D _\mu
A^a-i(q_L^a)^\dagger (\bar{\sigma }^\mu D_\mu )^{ab}q_L ^b
\nonumber\\&&
+(F^a)^* F^a-i(\lambda ^a
_R)^\dagger (\sigma ^\mu D_\mu)^{ab} \lambda
_R^b
\nonumber\\&&
+g\sqrt{2}\epsilon ^{abc}(A^a)^* (\lambda_R^b)^\dagger
 q_L^c-g\sqrt{2}\epsilon ^{abc}(q_L^a)^\dagger \lambda _R^b A^c
\nonumber\\&&
-\frac{1}{4}F^a_{\mu\nu }F^{a\mu \nu }+\frac 12
(D^a)^2
-ig\epsilon ^{abc}D^a(A^b)^* A^c.\label{LugWZSYM2}
\end{eqnarray}
Here $\lambda ^a$ and $q^a$ are the the two spinor fields of the theory, $A^a$
is the complex scalar field and $A^a_\mu $ the gauge field with the
corresponding field strength $F^a_{\mu \nu }$, while $D^a$ and $F^a$ are
auxiliary fields. As in \cite{GHN}  the conventions of Wess and Bagger
\cite{WB} are used, with $\eta ^{\mu \nu }=(-1,1,1,1)$, while $\sigma
^\mu=(-1,\vec{\sigma}),\hspace{1 mm} 
\bar{\sigma } ^\mu=(-1,-\vec{\sigma})$, where $\vec{\sigma}$
are the Pauli matrices. $\epsilon ^{abc}$ is the standard
Levi-Civita symbol.

The flat directions, where 
gauge invariance is spontaneously broken, are now
characterized by
\begin{equation}
D^a=0;\hspace{1 mm}A^a=(A^a)^* \neq 0.
\label{flatflat}
\end{equation}

Continuation of the fermionic part of the
action to Euclidean space requires some care because the
spinor fields, when expressed as Dirac spinors, obey the
Majorana condition. It is carried out by the
Vainshtein-Zakharov doubling trick
\cite{Vainshtein}. We skip the detailed manipulations, which
run as in \cite{GHN}. The outcome is the fermionic
Lagrangian:
\begin{eqnarray}
&&L_{N=2, {\rm Fermi,\hspace{1 mm}Euclid}}
\nonumber\\&&
=
-(\lambda ^a_A)^\dagger(\sigma \cdot
D )^{ab}\lambda ^b_B-(\lambda ^a_B)^\dagger(\bar{\sigma }\cdot
D )^{ab}\lambda ^b_A
\nonumber \\&&
-(q_A^a)^\dagger(\sigma \cdot D ^{ab}q_B^b
-(q_B^a)^\dagger(\bar{\sigma  }\cdot D^{ab} q_A^b
\nonumber \\&&
+g\sqrt{2}\epsilon ^{abc}(A^a)^* (\lambda_B^b)^\dagger
 q_B^c-g\sqrt{2}\epsilon ^{abc}(q_A^a)^\dagger \lambda _A^b A^c
\nonumber\\&&
-g\sqrt{2}\epsilon ^{abc}(A^a)^* q_B^\dagger
 \lambda _B^c+g\sqrt{2}\epsilon ^{abc}
(\lambda _A^a)^\dagger q_A^b A^c 
\label{lagslut}
\end{eqnarray}
where now $\bar{\sigma }_\mu =(i\vec{\sigma}, 1),\hspace{1 mm}\sigma _\mu
=(-i\vec{\sigma}, 1)$, and where two different
Weyl spinors are present, labelled $A$ and $B$, for each
Minkowski space Weyl spinor.

\section{Instanton mass corrections}
\setcounter{equation}{0}

\subsection{General setting}

Ignoring fermions and picking a particular flat direction, one
fulfils (\ref{flatflat}) by having
\begin{equation}
A^3=(A^3)^*=A\neq 0 
\label{Insatz}
\end{equation}
and  $A^1,A^2$ vanish.
Then the gauge field equations are:
\begin{equation}
\partial _\mu F^a_{\mu \nu }+g\epsilon ^{ab}(A^b_\mu F^3_{\mu \nu }-A^3_\mu
F^b_{\mu \nu }) -2g^2A^2A_\nu^a=0;\hspace{1 mm}a=1,2
\label{geqqo}
\end{equation}
and
\begin{eqnarray}&&
\partial _\mu F^3_{\mu \nu }+g\epsilon ^{ab}A^a_\mu F^b _{\mu \nu }
=0
\label{griqua}
\end{eqnarray}
while the scalar field equations are:
\begin{equation}
\partial ^2A-g^2A^a_\mu A^a_\mu A=0;\hspace{1 mm}a=1,2
\label{quodquod}
\end{equation}
and
\begin{equation}
2g\epsilon ^{ab}A^b_\mu \partial _\mu A+g^2A^a_\mu A^3_\mu A=0;\hspace{1
mm}a,b=1,2. 
\label{qeeego}
\end{equation}

The gauge fields are assumed to  have the following form in the singular
gauge:
\begin{equation}
A^a_{\mu }
=-\frac{1}{g}\bar{\eta }^a_{\mu \nu }\partial _{\nu }\log
\alpha _\perp;\hspace{1 mm}a=1,2 , \hspace{1 mm} A^3_{\mu }
=-\frac{1}{g}\bar{\eta }^3_{\mu \nu }\partial _{\nu }\log
\alpha _\parallel ,
\label{unsatz}
\end{equation}
where $\bar{\eta }^a_{\mu \nu }$ is the standard 't Hooft
symbol. Here the prepotentials
$\alpha _\perp$ and
$\alpha _\parallel$, as well as the functioon $A$ introduced
in (\ref{Insatz}), are real  functions of the parameter
$t=\frac{\rho^2}{x^2}$, with  $\rho $ the scale of
the instanton.
Then (\ref{qeeego}) is trivially valid.

Expressed in terms of the prepotentials, the field strength components are
\begin{eqnarray}&&
F^a_{\mu \nu }
=-\frac{4}{g}\bar{\eta }^a_{\mu\nu
}\frac{ t^2}{\rho^2}\frac{d\log\alpha
_\perp}{dt}(1-t\frac{d\log \alpha
_\parallel}{dt})
\nonumber\\&&
-\frac{4}{g}(\bar{\eta }^a_{\nu \lambda}x_\mu
x_\lambda -\bar{\eta }^a_{\mu \lambda}x_\nu
x_\lambda)(\frac{t^2}{\rho^4}\frac{d}{dt}t^2
\frac{d\log \alpha _\perp}{dt}
\nonumber\\&&
-\frac{t^4}{\rho ^4}\frac{d\log \alpha
_\perp}{dt}\frac{d\log\alpha
_\parallel}{dt});\hspace{1 mm}a=1,2
\label{kadabra}
\end{eqnarray}
and
\begin{eqnarray}&&
F^3_{\mu \nu }=-\frac{4}{g}\bar{\eta }^3_{\mu\nu
}\frac{ t^2}{\rho^2}(\frac{d\log \alpha
_\parallel}{dt}-t(\frac{d\log \alpha
_\perp}{dt})^2)
\nonumber\\&&-\frac{4}{g}(\bar{\eta }^3_{\nu \lambda}x_\mu
x_\lambda -\bar{\eta }^3_{\mu \lambda}x_\nu
x_\lambda)(\frac{t^2}{\rho^4}\frac{d}{dt}t^2
\frac{d\log \alpha _\parallel}{dt}
\nonumber\\&&-\frac{t^4}{\rho ^4}(\frac{d\log
\alpha _\perp}{dt})^2)
\label{meadowroad}
\end{eqnarray}
where  the following identity was used:
\begin{equation}
\epsilon ^{abc}\bar{\eta }^b_{\mu \lambda }\bar{\eta }^c_{\nu \rho
}x_\lambda x_\rho=x^2\bar{\eta }^a_{\mu \nu }+\bar{\eta }^a_{\nu \lambda
}x_\mu x_\lambda-\bar{\eta }^a_{\mu \lambda
}x_\nu x_\lambda .
\end{equation}

The action density is then (for a flat direction): 
\begin{eqnarray}&&
- (D_\mu A^a)^\dagger D_\mu A^a-\frac 14F^a_{\mu \nu
}F^a_{\mu
\nu }
\nonumber\\&&
=-\frac{4t^2}{\rho
^2}(\frac{dA}{dt})^2-\frac{8t^3}{\rho
^2}(\frac{d\log \alpha
_{\perp}}{dt})^2A^2
\nonumber\\&&
-\frac{16}{g^2}\frac{t^4}{\rho^4}(\frac{d\log\alpha _\perp
}{dt}(t\frac{d\log\alpha _\parallel}{dt}-1))^2
\nonumber\\&&
-\frac{8}{g^2}\frac{t^4}{\rho^4}(t(\frac{d\log\alpha _\perp}{dt})^2
-\frac{d\log\alpha
_\parallel}{dt})^2
\nonumber\\&&-\frac{16}{g^2}\frac{t^4}{\rho^4}(\frac{d}{dt}t
\frac{d\log \alpha  _\perp}{dt})^2
%\nonumber\\&&
-\frac{8}{g^2}\frac{t^4}{\rho^4}(\frac{d}{dt}t
\frac{d\log \alpha_\parallel}{dt})^2.
\label{Faidman}
\end{eqnarray}
(\ref{Faidman}) is negative semidefinite, and hence each
term has to give a finite contribution to the action. From
this observation follows  bonds on the functions 
$\alpha _{\perp}$, $\alpha _{\parallel}$ and $A$. The analysis
runs as in
\cite{NN}, with the difference that only $\alpha _\perp$ falls off
exponentially at small $t$ whereas 
$\alpha _\parallel$ and $A$ have a power law decrease. The
outcome is that $\alpha _\perp $ and $\alpha _\parallel $ at
large $t$ at most should grow logarithmically and for small
$t$ the leading terms of $\alpha _\perp$ should  conspire to
the modified Bessel funtion $K_1$:
\begin{equation}
\alpha_\perp=\alpha_{\perp, 0}+\alpha_{\perp,
2}+\alpha_{\perp,4}+\cdots\simeq
1+\sqrt{2t}\rho gv K_1(\sqrt {\frac 2t}\rho gv),
\label{an1}
\end{equation}
thus ensuring exponential
falloff of the gauge field in the nonflat directions.

The field equations (\ref{geqqo}), (\ref{griqua}) and
(\ref{quodquod}) are in terms of the functions $\alpha _\perp$,
$\alpha _\parallel
$ and $A$:
\begin{eqnarray}&&
\alpha_\perp^2\frac{d}{dt}\left(\alpha
_\perp^{-3}t^3\frac{d^2\alpha _\perp}{dt^2}\right)-3 t^2\alpha
_\perp ^{-1}\frac{d\alpha _\perp}{dt}(\alpha
_\perp^{-1}\frac{d\alpha
_\perp}{dt}-\alpha
_\parallel^{-1}\frac{d\alpha
_\parallel }{dt})
\nonumber\\&&
-t^3\alpha_\perp^{-1}\frac{d\alpha _\perp
}{dt}(\alpha _\parallel^{-2}(\frac{d\alpha_\parallel}{dt})^2-\alpha
_\perp^{-2}(\frac{d\alpha_\perp}{dt})^2)
\nonumber\\&&
=\frac{\rho^2g^2A^2}{2 }\alpha_\perp^{-1}
\frac{d\alpha_\perp}{dt},
\label{ali1}
\end{eqnarray}
\begin{eqnarray}&&
\alpha_\parallel^2\frac{d}{dt}\left(\alpha
_\parallel^{-3}t^3\frac{d^2\alpha
_\parallel}{dt^2}\right)
+3 t^2(\alpha _\perp ^{-2}(\frac{d\alpha
_\perp}{dt})^2-\alpha
_\parallel^{-2}(\frac{d\alpha
_\parallel }{dt})^2)
\nonumber\\&&
+2t^3\alpha_\parallel^{-1}\frac{d\alpha _\parallel
}{dt}(\alpha _\parallel^{-2}(\frac{d\alpha_\parallel}{dt})^2-\alpha
_\perp^{-2}(\frac{d\alpha_\perp}{dt})^2)
=0
\label{ali2}
\end{eqnarray}
and
\begin{eqnarray}
\frac{d^2A}{dt^2}-\frac{2}{\alpha
_\perp^2}\left(\frac{d\alpha _\perp}{dt}\right)^2A
  =0.  \label{ali3}
\end{eqnarray}

%%%%%%%%%%%%%%%%%%%%%%%%%%%%
\subsection{Iteration of the bosonic field equations up to
second order}
\label{sqcditeration}
%%%%%%%%%%%%%%%%%%%%%%%%%%

(\ref{ali1}), (\ref{ali2})
and (\ref{ali3}) are solved by iteration, where in the two
lowest orders:
\begin{equation}
\alpha_{\perp, 0}=\alpha_{\parallel,
0}=1+t;\hspace{1 mm}
A_1=\frac{v}{1+t}
\label{virajpet}
\end{equation}
with $v$ a free parameter, in terms of which the iteration
is carried out, and the subscript indicates the order.

At second order we require, according to the discussion
leading to (\ref{an1}):
\begin{equation}
\alpha _{\perp, 2} \simeq \alpha
_{\perp,2,{\rm min}}
\label{alfafato}
\end{equation}
near $t\simeq 0$, with
\begin{equation}
\alpha
_{\perp,2,{\rm min}}=\frac{\rho^2g^2v^2}{2}(\log
\frac{\rho^2g^2v^2}{2t}+2\gamma-1)
\label{minimal}
\end{equation}
by the power series expansion of the modified Bessel
function $K_1$.

(\ref{ali1}) and (\ref{ali2}) are at second order:
\begin{eqnarray}&&
(1+t)^2\frac{d}{dt}\left((\frac{t}{1+t})^3
\frac{d^2\alpha _{\perp,2}}{dt^2}\right)
-\frac{t^2}{1+t}(3-\frac{2t}{1+t})\frac{d}{dt}\frac{\alpha
_{-,2}}{1+t}
\nonumber\\&&
=\frac{\rho^2g^2v^2}{2 }\frac{1}{(1+t)^3}
\label{aly1}
\end{eqnarray}
and
\begin{equation}
%&&
(1+t)^2\frac{d}{dt}\left((\frac{t}{1+t})^3
\frac{d^2\alpha _{\parallel,2}}{dt^2}\right)
%\nonumber\\&&
+\frac{t^2}{1+t}(6-\frac{4t}{1+t})\frac{d}{dt}\frac{\alpha
_{-,2}}{1+t}=0
\label{aly2}
\end{equation}
with
\begin{equation}
\alpha _{-,2}=\alpha
_{\perp,2}-\alpha _{\parallel,2}.
\end{equation}

The terms involving $\alpha _{-,2}$
are eliminated by forming a linear combination of
(\ref{aly1}) and (\ref{aly2}), leading to
\begin{equation}
(1+t)^2\frac{d}{dt}\left((\frac{t}{1+t})^3
\frac{d^2(2\alpha _{\perp,2}+\alpha
_{\parallel,2})}{dt^2}\right) =\frac{\rho^2g^2v^2}{2
}\frac{2}{(1+t)^3}
\end{equation}
whence 
\begin{equation}
\frac{d^2(2\alpha _{\perp,2}+\alpha _{\parallel,2})}{dt^2}
=-\frac{\rho^2g^2v^2}{4
}\frac{1}{t^3(1+t)}+c_{2\perp+\parallel:2}(\frac{1+t}{t})^3
\label{torquilston}
\end{equation}
where $c_{2\perp+\parallel:2}$ is an integration constant that
should be chosen such that the leading singularity for
$t\rightarrow 0$ is eliminated, i.e.
\begin{equation}
c_{2\perp+\parallel:2}=\frac{\rho ^2g^2v^2}{4}.
\label{ulriqa}
\end{equation}
However, a nonvanishing value of $c_{2\perp+\parallel:2}$ 
introduces on the right hand side of (\ref{torquilston}) terms
that for $t\rightarrow
\infty$ are nonvanishing or only fall off as $\frac
1t$. These terms must be eliminated by a constraint. Hence,
(\ref{torquilston}) has to be replaced by:
\begin{equation}
\frac{d^2(2\alpha _{\perp,2}+\alpha _{\parallel,2})}{dt^2}
=\frac{\rho^2g^2v^2}{4
}(-\frac{1}{t^3(1+t)}+\frac{1}{t^3}+\frac{3}{t^2})
\label{frontdeboeuf}
\end{equation}
that serves as a guideline for the detailed determination of
possible constraints.

From (\ref{aly1})-(\ref{aly2}) also follows 
\begin{eqnarray}&&
(1+t)^2\frac{d}{dt}\left((\frac{t}{1+t})^3
\frac{d^2\alpha _{-,2}}{dt^2}\right)-
\frac{t^2}{1+t}(9-\frac{6t}{1+t})
\frac{d}{dt}\frac{\alpha_{-,2}}{1+t}
\nonumber\\&&
=\frac{\rho^2g^2v^2}{2 }\frac{1}{(1+t)^3}.
\label{aslak1}
\end{eqnarray}
This equation can be solved by quadrature, but it is
convenient instead to reformulate it slightly.

The function $\Psi _2$, defined by:
\begin{equation}
\Psi _2=\frac{1}{1+t}\frac{d}{dt}\frac{\alpha
_{-,2}}{1+t}.
\label{petšfi}
\end{equation}
fulfils in terms of the variable $u=\frac{t}{1+t}$ 
an inhomogeneous hypergeometric equation:
\begin{equation}
u(1-u)\frac{d^2\Psi _2(u)}{du^2}+3\frac{d\Psi
_2(u)}{du}=\frac{\rho ^2g^2v^2}{2}\frac{(1-u)^4}{u^2}.
\label{tutmoses}
\end{equation}
The general version (\ref{tutankhamon}) of this equation is
treated in detail in the appendix, and  the general solution
is given in (\ref{weeweewee}).
 In second order 
the solution is, with the integration constant chosen to
make $\Psi_2(1)=0$:
\begin{equation}
\Psi_2(u)=\frac{\rho^2g^2v^2}{2}
(-\frac 1u-\frac
32+3u-\frac 12u^2-3\log u).
\label{copperscot}
\end{equation}
Inserting (\ref{copperscot}) into (\ref{petšfi}) one gets the
result
\begin{equation}
\frac{\alpha _{-,2}}{1+t}
=\frac{\rho^2g^2v^2}{2}(\frac 12\log u-\frac 32\frac{\log
u}{(1-u)^2}-\frac 32\frac{1}{1-u})\mid
_{u=\frac{t}{1+t}}
%\nonumber\\&&
+C_1
\label{atternat}
\end{equation}
with $C_1$ an  integration constant determined below.

The integrated
versions of (\ref{aly1}) and (\ref{aly2}) are:
\begin{eqnarray}&&
(\frac{t}{1+t})^3\frac{d^2\alpha
_{\perp,2}}{dt^2}
=-\int  _{u}^1((\frac{u'}{1-u'})^2(3-2u')\Psi_2(u')
\nonumber\\&&
+\frac{\rho ^2g^2v^2}{2}(1-u')^3)du'\mid _{u=\frac{t}{1+t}}
+c_{\perp;2}
\label{alric}
\end{eqnarray}
and
\begin{equation}
(\frac{t}{1+t})^3\frac{d^2\alpha
_{\parallel,2}}{dt^2}=\int
_{u}^1(\frac{u'}{1-u'})^2(6-4u')\Psi_2(u')du'\mid
_{u=\frac{t}{1+t}}
%\nonumber\\&&
+c_{\parallel;2}.
\label{olric}
\end{equation}
Here $c_{\perp;2}$ and $c_{\parallel;2}$ are integration 
constants. Since $\Psi _2(u)=O((1-u)^4)$  for
$u\rightarrow 1$  it follows that the
integrals in (\ref{alric}) and (\ref{olric}) are
$O((1-u)^3)=O(t^{-3})$ in this limit that after two
integrations are
$O(t^{-1})$. Nonzero values of the integration
constants, however, lead to singular solutions for
$t\rightarrow \infty$.

Considering separately
\begin{equation}
(\frac{t}{1+t})^3\frac{d^2\alpha
_{\perp,2}}{dt^2}\simeq c_{\perp;2}
\label{kualalumpur}
\end{equation}
and
\begin{equation}
(\frac{t}{1+t})^3\frac{d^2\alpha
_{\parallel,2}}{dt^2}\simeq c_{\parallel;2}
\label{kilimanjaro}
\end{equation}
one obtains the solutions
\begin{equation}
\alpha _{\perp,2}\simeq c_{\perp;2}(\frac{1}{2t}-3\log t+3(t\log
t-t)+\frac 12 t^2)
\label{kimberley}
\end{equation}
and
\begin{equation}
\alpha _{\parallel,2}\simeq c_{\parallel;2}(\frac{1}{2t}-3\log
t+3(t\log t-t)+\frac 12 t^2)
\label{equator}
\end{equation}
where terms involving $t\log t$ and $t^2$ must be
eliminated by a constraint.  This is achieved by making in
(\ref{kualalumpur}) and (\ref{kilimanjaro}) the replacements:
\begin{eqnarray}&&
c_{\perp;2}\rightarrow
c_{\perp;2}-c_{\perp;2}(\frac{t}{1+t})^3(1+\frac
3t);\nonumber\\&&c_{\parallel;2}\rightarrow
c_{\parallel;2}-c_{\parallel;2}(\frac{t}{1+t})^3(1+\frac 3t).
\label{replac}
\end{eqnarray}

The integral occurring in
(\ref{alric}) and (\ref{olric})  is by insertion of
 (\ref{copperscot}) explicitly:
\begin{eqnarray}&&
\int _u^1(\frac{u'}{1-u'})^2(3-2u')\Psi_2(u')du'
\nonumber\\&&
=\frac{\rho
^2g^2v^2}{2} \Bigg(-\frac{1}{12}
+u^3(\frac{3\log u}{1-u}+\frac
32\frac 1u+\frac{11}{6}-\frac 14u)\Bigg).
\end{eqnarray}
The requirement of acceptable behaviour of $\alpha _{\perp,2}$
and $\alpha _{\parallel, 2}$ near
$t\simeq 0$ fixes the two integration constants:
\begin{equation}
c_{\perp;2}=c_{\parallel;2}=\frac{\rho ^2g^2v^2}{12}
\label{womaninwhite}
\end{equation}
thus making the right hand sides of (\ref{alric}) and
(\ref{olric}), modified according to (\ref{replac}), vanish
for
$u\rightarrow 0$ ($t\rightarrow 0$). There is obviously
agreement between (\ref{ulriqa}) and (\ref{womaninwhite}),
with $c_{2\perp+\parallel:2}=2c_{\perp;2}+c_{\parallel;2}$.

 (\ref{alric}) becomes after these transformations:
\begin{equation}
\frac{d^2\alpha_{\perp,2}}{dt^2}
=\frac{\rho^2g^2v^2}{2}
(3(1+t)\log \frac{1+t}{t}-3-\frac{3}{2}\frac{1}{t}
+\frac{1}{t^2}).
\label{casuar} 
\end{equation}
$\alpha _{\perp,2}$ is allowed to grow only
logarithmically for $t\rightarrow 0$, where  it has to agree with (\ref{alfafato}). This fixes the
integration constants of (\ref{casuar}) completely and the
solution is:
\begin{eqnarray}&&
\alpha _{\perp,2}=\frac{\rho^2g^2v^2}{4}((1+t)^3\log \frac
{1+t}{t}+\log t-\frac 52t-t^2)
\nonumber\\&&+\alpha
_{\perp,2,{\rm min}}.
\label{xhuxha}
\end{eqnarray}

Likewise it follows from (\ref{olric})
\begin{equation}
\frac{d^2\alpha_{\parallel,2}}{dt^2}=\frac{\rho^2g^2v^2}{2}
(-6(1+t)\log \frac{1+t}{t}+6+\frac 52\frac {1}{t}+\frac 12
\frac{1}{1+t})
\label{elkj¾rsultan}
\end{equation}
that with (\ref{casuar}) is
consistent with
(\ref{frontdeboeuf}). From
(\ref{elkj¾rsultan}) follows by integration and taking the
boundary conditions into account:
\begin{eqnarray}&&
\alpha_{\parallel,2}=\frac{\rho^2g^2v^2}{2}
(-(1+t)^3\log \frac{1+t}{t}+\frac
52t+t^2)
\nonumber\\&&
+\frac{\rho^2g^2v^2}{4}((1+t)\log \frac
{1+t}{t}+\log t)
\nonumber\\&&
+\frac{\rho^2g^2v^2}{2}(\log
\frac{\rho^2g^2v^2}{2t}+2\gamma-1)+C_2
\label{mimseyvane}
\end{eqnarray}
where the terms in the first line  have logarithmic growth
for $t\rightarrow
\infty $ and where $C_2$ is an integration constant.
From (\ref{xhuxha}) and (\ref{mimseyvane}) is formed
$\alpha_{-,2}=\alpha_{\perp,2}-\alpha_{\parallel,2}$, and the outcome is
in agreement with (\ref{atternat}) for
\begin{equation}
C_1=-\frac{\rho^2g^2v^2}{2}\frac {3}{4},\hspace{1 mm}
C_2=\frac{\rho^2g^2v^2}{2}\frac {9}{4}.
\label{kamehameha}
\end{equation}

The modification of the field equations means that in
(\ref{ali1}) and (\ref{ali2}) an extra term occurs on the
right hand side:
\begin{equation}
-\frac{\rho ^2g^2v^2}{2}\frac{t}{(1+t)^2}.
\label{elkj¾rbusk}
\end{equation}

So far the modification of the field equations,
apparent in the replacement (\ref{replac}) in (\ref{alric})
and (\ref{olric}), with the two integration constants given
by (\ref{womaninwhite}), has been as small as possible in the
sense that it is the modification that is necessary in order
to obtain permitted asymptotic behaviour of the instanton
solution. However, once this objective is achieved,
additional constraint terms that do not upset the
asymptotic behaviour can be added freely to the right
hand sides of (\ref{alric}) and (\ref{olric}). One
might thus modify (\ref{alric}) in such a way that
$\alpha _{\perp,2}$ reduces to $\alpha _{\perp,2
,{\rm min}}.$ Then the right hand side of (\ref{alric})
should contain the additional term
\begin{equation}
-3(\frac{t}{1+t})^3\frac{\rho ^2g^2v^2}{2}((1+t)\log
\frac{1+t}{t}-1-\frac{1}{2}\frac 1t).
\label{gurthcedric}
\end{equation}
In order to keep the constraint gauge covariant one then should
add this term also to the right hand side of (\ref{olric}).

It is preferable in general to restrict additional
constraint terms by gauge covariance, having the same
constraint terms in all modifications of (\ref{alric})
and (\ref{olric}) as we have by (\ref{womaninwhite}) in the simplest case
considered here. Then
$\alpha _{-,2}$ given in (\ref{atternat}) (with
(\ref{kamehameha})) is independent of the choice of
constraint.

\subsection{Third order and fourth order}
 
The third order  scalar field  is
according to (\ref{ali3}) determined by:
\begin{equation}
(1+t)^2\frac{d^2A_3}{dt^2}-2A_3
 =4v\frac{d}{dt}\frac{\alpha _{\perp,2}}{1+t}
\end{equation}
whence
\begin{equation}
A_3=-\frac{v\alpha _{\perp, 2}(t)}{(1+t)^2}-v(1+t)^2\int
_t^\infty
\frac{d\alpha _{\perp, 2}(t')}{dt'}\frac{dt'}{(1+t')^4}.
\end{equation}
Here  an integration constant is taken equal to zero to
ensure  that $A_3$ is
$O(t^{-2})$ for $t\rightarrow \infty $, since
$\frac{d\alpha _{\perp, 2}}{dt}$ is $O(t^{-1})$ in this
limit because of the constraint. This means that no additional
constraint is necessary here. 
 Using (\ref{xhuxha}) one   obtains the final result: 
\begin{eqnarray}&&
A_{3}=-\frac{v\alpha
_{\perp,2}}{(1+t)^2}+(1+t)^2\frac{\rho ^2g^2 v ^3}{4}
   ((\frac{3t}{1+t}+2)\log \frac{1+t}{t}
\nonumber\\&&
-\frac{5}{1+t}
+\frac{1}{2}\frac{1}{(1+t)^2}-\frac{1}{6}\frac{1}{(1+t)^3}).
\label{tildeftre}
\end{eqnarray}

Replacing $\alpha _{\perp,2}$ with $\alpha
_{\perp,2,{\rm min}}$ in (\ref{minimal}), which as
mentioned above can be obtained by a modified
constraint, we get instead of (\ref{tildeftre}):
\begin{eqnarray}&&
A_3
=-\frac{v\alpha_{\perp,2}(t)}{(1+t)^2}
+(1+t)^2\frac{\rho
^2g^2v^3}{2}(\log
\frac{1+t}{t}-\frac{1}{1+t}
\nonumber\\&&
-\frac{1}{2}\frac{1}{(1+t)^2}
-\frac{1}{3}\frac{1}{(1+t)^3}).
\label{yaguarundi}
\end{eqnarray}

The equations determing the fourth order gauge
prepotentials are according to
(\ref{ali1}) and (\ref{ali2}):
\begin{eqnarray}&&
(1+t)^2\frac{d}{dt}\left((\frac{t}{1+t})^3
\frac{d^2\alpha _{\perp,4}}{dt^2}\right)
-\frac{t^2}{1+t}(3-\frac{2t}{1+t})\frac{d}{dt}\frac{\alpha
_{-,4}}{1+t}
\nonumber\\&&
=3(1+t)^2\frac{d}{dt}\Bigg(\alpha_{\perp,
2}\frac{t^3}{(1+t)^4}\frac{d^2\alpha_{\perp,2}}{dt^2}\Bigg)
\nonumber\\&&
+\frac{\rho^2g^2v^2}{2
}(1+t)^2\frac{d}{dt}\frac{\alpha
_{\perp,2}}{(1+t)^5}+\tilde{\chi} _{\perp, 4}
\label{aly3}
\end{eqnarray}
and
\begin{eqnarray}&&
(1+t)^2\frac{d}{dt}\left((\frac{t}{1+t})^3
\frac{d^2\alpha _{\parallel,4}}{dt^2}\right)
+\frac{t^2}{1+t}(6-\frac{4t}{1+t})\frac{d}{dt}\frac{\alpha
_{-,4}}{1+t}
\nonumber\\&&
=3(1+t)^2\frac{d}{dt}\Bigg(\alpha_{\parallel,
2}\frac{t^3}{(1+t)^4}\frac{d^2\alpha_{\parallel,2}}{dt^2}\Bigg)
+\tilde{\chi } _{\parallel, 4}
\label{aly4}
\end{eqnarray}
with
\begin{eqnarray}&&
\tilde{\chi } _{\perp, 4}=
-\frac{3t^2}{(1+t)^2}(\alpha _{\perp,
2}\frac{d}{dt}\frac{\alpha _{\perp, 2}}{1+t}-\alpha _{\parallel,
2}\frac{d}{dt}\frac{\alpha _{\parallel, 2}}{1+t})
\nonumber\\&&
-\frac{t^3}{1+t}
\Bigg(\frac{d}{dt}\frac{\alpha _{\perp
,2}}{1+t}(\frac{d}{dt}\frac{\alpha _{\perp
,2}}{1+t}-2\frac{\alpha _{\perp,2}}{(1+t)^2})
\nonumber\\&&
-\frac{d}{dt}\frac{\alpha
_{\parallel,2}}{1+t}(\frac{d}{dt}\frac{\alpha
_{\parallel,2}}{1+t}-2\frac{\alpha
_{\parallel,2}}{(1+t)^2})\Bigg)
\nonumber\\&&
+t^2(3+t)(\frac{d}{dt}\frac{\alpha_{\perp,2}}{1+t}-\frac{2\alpha_{\perp,2}}
{(1+t)^2})\Psi _2
\nonumber\\&&
 +\frac{\rho^2g^2v^2}{2
}\frac{2(\frac{{A}_3}{v}+\frac{\alpha _{\perp,2}}{(1+t)^2})}{(1+t)^2}
\label{chatter}
\end{eqnarray}
and
\begin{eqnarray}&&
\tilde{\chi } _{\parallel, 4}=-t^2
(3-\frac{2t}{1+t})\Bigg(\frac{d}{dt}\frac{\alpha _{\perp
,2}}{1+t}(\frac{d}{dt}\frac{\alpha _{\perp
,2}}{1+t}-2\frac{\alpha _{\perp,2}}{(1+t)^2})
\nonumber\\&&
-\frac{d}{dt}\frac{\alpha
_{\parallel,2}}{1+t}(\frac{d}{dt}\frac{\alpha
_{\parallel,2}}{1+t}-2\frac{\alpha
_{\parallel,2}}{(1+t)^2})\Bigg)
\nonumber\\&&
+(\frac{2t^2}{1+t}(6-\frac{4t}{1+t})\alpha
_{\parallel, 2}+4t^3\frac{d}{dt}\frac{\alpha
_{\parallel, 2}}{1+t})\Psi _2.
\label{cherusker}
\end{eqnarray}

For $t\rightarrow \infty$ the right-hand sides of
(\ref{aly3}) and (\ref{aly4}) are $O(t^{-2})$. This should be
compared to the right-hand side of the
second order equation (\ref{aly1}), which is $O(t^{-3})$ in
this limit.

The function $\Psi _4=\frac{1}{1+t}\frac{d}{dt}\frac{\alpha _{-,4}}{1+t}$ is
by (\ref{aly3})-(\ref{aly4}) a solution of the inhomogeneous hypergeometric
equation (\ref{tutankhamon}) with $X=X_4$ given by:
\begin{eqnarray}&&
\frac{t^2}{1+t} X_4 
=3(1+t)^2\frac{d}{dt}\Bigg(\frac{t^3}{(1+t)^4}(\alpha_{\perp,
2}\frac{d^2\alpha_{\perp,2}}
{dt^2}
\nonumber\\&&-\alpha_{\parallel,
2}\frac{d^2\alpha_{\parallel,2}}{dt^2})\Bigg)
+\frac{\rho^2g^2v^2}{2}(1+t)^2\frac{d}{dt}\frac{\alpha
_{\perp,2}}{(1+t)^3}
\nonumber\\&&
+\tilde{\chi } _{\perp, 4}-\tilde{\chi }_{\parallel, 4}.
\label{¿rnsager}
\end{eqnarray}
Most of the right-hand side of (\ref{¿rnsager}) is $O(t^0)$ in
the limit $t\rightarrow 0$ where
\begin{equation}
X_4\simeq -(\frac{\rho ^2g^2v^2}{2})^2\frac{1}{t^3}
\label{h¿jvig}
\end{equation}
while $X_4$ is $O(
t^{-3})$ for $t\rightarrow \infty $.

With $\Psi _4$ given by (\ref{weeweewee}) specialized to fourth order, its
derivative  has by (\ref{johannkepler}) the following leading
term for $u\rightarrow 0$ by an appropriate choice of the integration
constant:
\begin{equation}
\frac{d\Psi _4(u)}{du}\simeq\frac{1 }{u^3}\int
_{0}^u(u')^2(1-u')^{-4}X_4(u')du'.
\end{equation}
From (\ref{h¿jvig}) follows the splitting:
\begin{equation}
X_4(u)=\frac{\rho
^2g^2v^2}{2}\frac{(1-u)^4}{u^2}\frac{d\alpha
_{\perp,2}}{du}+\tilde{X}_4(u)
\end{equation}
where $\tilde{X}_4(u)=O(u^{-2})$ for $u\rightarrow 0$, and
hence:
\begin{equation}
\frac{d\Psi _4(u)}{du}\simeq\frac{1}{u^3}\frac{\rho ^2g^2v^2}{2}\alpha
_{\perp,2}(u)
\label{valdal}
\end{equation}
giving the leading singularity of $\Psi _4$ for $u\rightarrow 0$,
while
$\Psi _4 =O((1-u)^{4})$ near $u=1$ by the argument after
(\ref{weeweewee}).

By integration of
(\ref{aly3}) and (\ref{aly4}) follows:
\begin{eqnarray}&&
(\frac{t}{1+t})^3\frac{d^2\alpha
_{\perp,4}}{dt^2}=3\alpha_{\perp,
2}\frac{t^3}{(1+t)^4}\frac{d^2\alpha_{\perp,2}}{dt^2}+\frac{\rho
^2g^2v^2}{2}\frac{\alpha _{\perp,2}}{1+t}
\nonumber\\&&
-\int  _{u}^1
((\frac{u'}{1-u'})^2(3-2u')\Psi_4(u')
+\tilde{\chi}_{\perp,4}(u'))du'\mid
_{u=\frac{t}{1+t}}
\nonumber\\&&
+c_{\perp;4}
\label{alrick}
\end{eqnarray}
and
\begin{eqnarray}&&
(\frac{t}{1+t})^3\frac{d^2\alpha
_{\parallel,4}}{dt^2}=3\alpha_{\parallel,
2}\frac{t^3}{(1+t)^4}\frac{d^2\alpha_{\parallel,2}}{dt^2}
\nonumber\\&&
-\int
_{u}^1((\frac{u'}{1-u'})^2(-6+4u')\Psi_4(u')
+\tilde{\chi} _{\parallel,4}(u'))du')\mid
_{u=\frac{t}{1+t}}
\nonumber\\&&+c_{\parallel;4}
\label{olrick}
\end{eqnarray}
where the integrals as those of (\ref{alric})-(\ref{olric}) are
$O((1-u)^3)=O(t^{-3})$ for
$t\rightarrow \infty$.

The integration constants
$c_{\perp;4}$ and 
$c_{\parallel;4}$ are as at second order fixed by the requirement of
acceptable asymptotic behaviour of the gauge prepotentials for
$t\rightarrow 0$, to
\begin{equation}
c_{\perp;4}=\int  _{0}^1((\frac{u}{1-u})^2(3-2u)\Psi_4(u)
+\tilde{\chi }_{\perp, 4}(u))du
\label{albuquerque}
\end{equation}
and
\begin{equation}
c_{\parallel;4}=\int
_{0}^1((\frac{u}{1-u})^2(-6+4u)\Psi_4(u)+\tilde{\chi}_{\parallel,
4}(u))du.
\label{magalhaes}
\end{equation}
With this choice of integration constants it follows from
(\ref{alrick}) and (\ref{olrick}) for
$t\rightarrow 0$: 
\begin{equation}
\frac{d^2\alpha _{\perp,4}}{dt^2}\simeq \frac{\rho^2g^2v^2}{2
t^3}\alpha _{\perp,2}  
\label{greatexpectations} 
\end{equation}
and
\begin{equation}
\frac{d^2\alpha _{\parallel,4}}{dt^2}=O(t^{-2})
\end{equation}
and from (\ref{greatexpectations}):
\begin{equation}
\alpha _{\perp,4}\simeq \frac 12(\frac{\rho ^2g^2v
^2}{2})^2\left(\log\frac{\rho ^2 g^2v
^2}{2t}+2\gamma-\frac{5}{2}\right)\frac{1}{t}
\label{gyldenkrone}
\end{equation}
in agreement with (\ref{an1}). No other
terms singular as $\frac 1t$ for
$t\rightarrow 0$ occur in
$\alpha _{\perp, 4}$ or $\alpha _{\parallel, 4}$; this is the
reason why the integration constants $c_{\perp, 4}$ and
$c_{\parallel;4}$ have been chosen according to
(\ref{albuquerque}) and (\ref{magalhaes}), respectively.

The integrals (\ref{albuquerque}) and (\ref{magalhaes})
converge, since $\Psi _4(u)$ is $O((1-u)^4)$ for
$u\rightarrow 1$ and $O(u^{-2})$ for
$u\rightarrow 0$. The integration constants are again equal (cf.
(\ref{womaninwhite})). This is seen by 
 use of (\ref{¿rnsager}), (\ref{valdal}) and (\ref{tutankhamon}):
\begin{eqnarray}&&
c_{\perp;4}-c_{\parallel;4}
\nonumber\\&&
=\int _0^1\frac{d}{du}(u^3(\frac{d\Psi
_4(u)}{du'}+\frac{3\Psi _4(u)}{1-u}
\nonumber\\&&
 -3u^3(1-u)(\alpha
_{\perp,2}\frac{d^2\alpha_{\perp,2}}{dt^2}-\alpha
_{\parallel,2}\frac{d^2\alpha_{\parallel,2}}{dt^2})(u)
\nonumber\\&&
-\frac{\rho^2g^2v^2}{2}(1-u)^5\alpha
_{\perp,2}(u))du
\nonumber\\&&
 = -\lim _{u\rightarrow 0}(u^3\frac{d\Psi
_4(u)}{du}-\frac{\rho ^2g^2v^2}{2}\alpha _2(u))=0.
\label{khatinkle}
\end{eqnarray}

 This result 
can be used to get a different expression for
$c_{\perp;4}=c_{\parallel;4}$ from
(\ref{albuquerque})-(\ref{magalhaes}):
\begin{equation}
c_{\perp;4}=
\frac 13\int _0^1(2\tilde{\chi }_{\perp, 4}(u')+\tilde{\chi
}_{\parallel, 4}(u'))du'
\label{chelinita}
\end{equation}
where, according to (\ref{chatter}) and
(\ref{cherusker}):
\begin{eqnarray}&&
2\tilde{\chi}_{\perp,4}+\tilde{\chi }_{\parallel,4}
=  -\frac{\Psi _2}{1-u}(\frac{3u^2}{(1-u)^3}\Psi_2
\nonumber\\&&
+\frac{\rho^2g^2v^2}{2}\frac{2u^2}{1-u}(3-2u)(\log u +\frac 1u-1))\mid
_{u=\frac{t}{1+t}}
\nonumber\\&&
+\frac{\rho^2g^2v^2}{2
}\frac{4(\frac{{A}_3}{v}
+\frac{\alpha _{\perp,2}}{(1+t)^2})}{(1+t)^2}
\label{elverskud}
\end{eqnarray}
where also (\ref{atternat}) and
(\ref{kamehameha}) were used.

Using the result (\ref{tildeftre}) for the scalar field $A$ 
one  obtains a term of (\ref{chelinita}):
\begin{equation}
\frac 13 \frac{\rho ^2g^2v^2}{2}\int _0^1 \frac{4(\frac
{A_3}{v}+\frac{\alpha _{\perp,2}}{(1+t)^2})}{(1+t)^2}\mid _{t=\frac{u}{1-u}}du
=\frac 14(\frac{\rho ^2g^2v^2}{2})^2.
\label{tuscarora}
\end{equation}
Using instead (\ref{yaguarundi}) it is replaced by 
\begin{equation}
\frac
13(\frac{\rho ^2g^2v^2}{2})^2.
\label{mescalero}
\end{equation}
The rest of (\ref{chelinita}) only depends on
$\alpha _{-,2}$ and is thus independent of the precise
form of the second order constraint, provided the same
constraint is used for
$\alpha _{\perp,2}$ and $\alpha _{\parallel,2}$, i.e.
the constraint is chosen gauge covariant. By insertion of (\ref{copperscot})
one obtains:
\begin{eqnarray}&&
\frac{2}{3}(\frac{\rho^2g^2v^2}{2})^2
(\frac 16+\frac{1}{18}-\frac{1}{16}). 
\end{eqnarray}
and thus the value of the two fourth order integration
constants is:
\begin{equation}
c_{\perp;4}=c_{\parallel;4}=(\frac{\rho ^2g^2v^2}{2})^2
(\left\{\begin{array}{c}
            \frac 14\\\frac 13
            \end{array}  \right\}    +\frac 23(\frac
16+\frac{1}{18}-\frac{1}{16})).
\end{equation}

Nonzero integration constants upset, like in second
order, the asymptotic behaviour for $t\rightarrow \infty$ and
necessitate a constraint that modifies (\ref{alrick}) 
 according to (cf.
(\ref{replac})):
\begin{equation}
c_{\perp;4}\rightarrow
c_{\perp;4}-c_{\perp;4}(\frac{t}{1+t})^3(1+\frac
3t)
\end{equation}
and the same for (\ref{olrick}). This in its turn means that
(\ref{ali1}) and (\ref{ali2}) must have an extra term on the
right hand sides (cf. (\ref{elkj¾rbusk})):
\begin{equation}
-6c_{\perp;4}\frac{t}{(1+t)^2}.
\label{elkj¾rf¿nix}
\end{equation}

\subsection{Higher orders; the short-distance limit}
\label{shortdist} At fifth order one gets from (\ref{ali3}):
\begin{eqnarray}
&&(1+t)^2\frac{d^2A_5}{dt^2}-2A_5
\nonumber\\&&
  =4v\frac{d}{dt}\frac{\alpha
_{\perp,4}}{1+t}+2v(1+t)(\frac{d}{dt}\frac{\alpha_{\perp, 2}}{1+t})^2
    -\frac{4v\alpha_{\perp, 2}}{1+t}\frac{d}{dt}\frac{\alpha
_{\perp, 2}}{1+t}
\nonumber\\&&
+4(1+t)A_3\frac{d}{dt}\frac{\alpha _{\perp,2}}{1+t}.\end{eqnarray}
Here the right-hand side is $O(t^{-2})$ for $t\rightarrow \infty$,
and hence $A_5$ is also $O(t^{-2})$ in
this limit.

At sixth order  (\ref{ali1}) and
(\ref{ali2}) imply corresponding to (\ref{aly1}) and
(\ref{aly2}), with $\alpha _{-,6}=\alpha _{\perp,6}-\alpha
_{\parallel, 6}$:
\begin{eqnarray}&&
(1+t)^2\frac{d}{dt}\left((\frac{t}{1+t})^3
\frac{d^2\alpha _{\perp,6}}{dt^2}\right)-
\frac{t^2}{1+t}(3-\frac{2t}{1+t})\frac{d}{dt}\frac{\alpha
_{-,6}}{1+t}
\nonumber\\&&
=\frac{\rho
^2g^2v^2}{2}(1+t)^2\frac{d}{dt}\frac{\alpha
_{\perp,4}}{(1+t)^3}+\tilde{\chi } _{\perp, 6}
%\nonumber\\&&
=\chi  _{\perp, 6}
\label{aly5}
\end{eqnarray}
and
\begin{eqnarray}&&
(1+t)^2\frac{d}{dt}\left((\frac{t}{1+t})^3
\frac{d^2\alpha _{\parallel,6}}{dt^2}\right)
+\frac{t^2}{1+t}(6-\frac{4t}{1+t})\frac{d}{dt}\frac{\alpha
_{-,6}}{1+t}
\nonumber\\&&
=\chi  _{\parallel, 6}
\label{aly6}
\end{eqnarray}
where $\tilde{\chi } _{\perp,6}$ and $\chi _{\parallel, 6}$ are
$O(t^{-1})$ for $t\rightarrow 0$ and $O(t^{-2})$ for
$t\rightarrow
\infty.$ Here 
$\Psi _6=\frac{1}{1+t}\frac {d}{dt}\frac{\alpha _{-,6}}{1+t}$
was introduced,
which is a solution of (\ref{tutankhamon}), with
$X_6$ given by:
\begin{eqnarray}&&
\frac{t^2}{1+t} X_6=\chi _{\perp, 6}-\chi _{\parallel, 6}.
\label{zinklar}
\end{eqnarray}
For $t\rightarrow 0$ it follows from (\ref{zolikon}) that
$\Psi _6 =O(t^{-3})$  while  it is $O(t^{-4})$ for
$t\rightarrow
\infty$ by (\ref{weeweewee}) as at second and fourth
order.

By integration of (\ref{aly5}) and (\ref{aly6}) follows: 
\begin{eqnarray}&&
(\frac{t}{1+t})^3\frac{d^2\alpha
_{\perp,6}}{dt^2}
=-\int  _{u}^1(
(\frac{u}{1-u'})^2(3-2u')\Psi_6(u')
\nonumber\\&&
+\chi_{\perp,6}(u'))du'\mid
_{u=\frac{t}{1+t}}
\label{alriccus}
\end{eqnarray}
and
\begin{eqnarray}&&
(\frac{t}{1+t})^3\frac{d^2\alpha
_{\parallel,6}}{dt^2}
=-\int _{u}^1((\frac{u'}{1-u'})^2(-6+4u')\Psi_6(u')
\nonumber\\&&
+\chi _{\parallel,6}(u'))du')\mid
_{u=\frac{t}{1+t}},
\label{olriccus}
\end{eqnarray}
respectively. The right-hand sides of (\ref{alriccus}) and
(\ref{olriccus}) are
$O(t^{-2})$ for $t\rightarrow \infty$, and the integration
constants, which upset this asymptotic behaviour in second and
fourth order, can here be chosen equal to zero; they are no
longer fixed by requiring Bessel function behaviour of $\alpha
_\perp$ because the Bessel function term now is
$O(t^{-3})$.  In the limit $t\rightarrow 0$ one gets
from (\ref{alriccus}), using also (\ref{aly5}) and
(\ref{gyldenkrone}):
\begin{equation}
\frac{d^2\alpha _{\perp,6}}{dt^2}\simeq  \frac 12(\frac{\rho
^2g^2v^2}{2})^3(\log \frac{\rho ^2g^2v^2}{2t}+2\gamma-\frac 52)
\frac{1}{t^4}
\end{equation}
with the solution
\begin{equation}
\alpha _{\perp,6}\simeq \frac{1}{12} (\frac{\rho
^2g^2v^2}{2})^3(\log \frac{\rho ^2g^2v^2}{2t}+2\gamma-\frac
{10}{3})
\frac{1}{t^2}
\end{equation}
in agreement with (\ref{an1}).

The iteration at higher orders proceeds in the same way.
It is proved by induction  that the $n$th order terms, $n\neq 0$,
of
$\alpha _{\perp}(t)$ and $\alpha _{\parallel}(t)$  as well
as
$t^2 A(t)$ in all orders at most have logarithmic growth for
$t\rightarrow \infty$. At each order the
complication arising from the coupling of the two equations
of the gauge preponentials is handled in the same way as in
the sixth order calculation, by means of (\ref{weeweewee}) and the asymptotic
estimates it implies.

\subsection{Constraint terms in the vector field
equation} It was found in (\ref{elkj¾rbusk}) and
(\ref{elkj¾rf¿nix}) that the equations (\ref{ali1}) and
(\ref{ali2})   both require the following
additional term on the right-hand side
\begin{equation}
-6c\frac{t}{(1+t)^2}
\end{equation}
with $c=c_{\perp;2}+c_{\perp;4}$.
Thus the gauge field equation has to be modified to:
\begin{eqnarray}&&
\partial _\mu F^a_{\mu \nu }+g\epsilon
^{abc}A^b_\mu F^c_{\mu
\nu }+\frac{c}{g}
 \bar{\eta}^a_{\nu
\lambda }x
_\lambda\frac{48\rho^2}{x^2   (\rho^2+x^2)^2}
\nonumber\\&&-2g^2(A^bA^bA_\nu^a-A^aA^bA_\nu^b)=0
\label{feegee}
\end{eqnarray}
where the extra term can be obtained from a source term in
the Lagrangian that arises from a constraint on the path
integral. This modification of the gauge field equation has
the same structure as that determined for the case
 where the scalar field is in the
fundamental representation \cite{NN}, \cite{GHN}. 

Modifying the equations according to (\ref{gurthcedric}), one obtains
correspondingly an additional source term   in the gauge field equation.

\subsection{Leading and subleading terms at
large distances}
The leading terms for $t\rightarrow 0$ fulfil in each order of
$v$:
\begin{equation}
\alpha _{\perp,\parallel,n}\propto t^{1-\frac
n2},\hspace{1 mm}A_n\propto t^{\frac 32-\frac
n2},\hspace{1 mm}n>1
\end{equation}
and
\begin{equation}
\alpha _{\perp,\parallel,0}\simeq 1,\hspace{1 mm}
A_1\simeq v.
\end{equation}
The sums of the leading terms are denoted $\alpha
_\perp^{(2)}$, $\alpha  _\parallel^{(2)}$ and $A^{(2)}$
where the superscript indicates that  they 
contain a factor
$\rho ^2$ when expressed in terms of $\rho $ and $x$ (when the functions are
expressed in terms of $\rho $ and $\sqrt t$, the superscript indicates the
combined power of $\rho $ and $\sqrt t$). Then 
(\ref{ali1}), (\ref{ali2}) and (\ref{ali3}) imply:
\begin{equation}
\frac{d}{dt}(t^3\frac{d^2\alpha
_\perp^{(2)}}{dt^2})=
\frac{\rho ^2g^2v^2}{2}\frac{d\alpha  _\perp^{(2)}}{dt},
\label{forethought}
\end{equation}
\begin{equation}
\frac{d}{dt}(t^3\frac{d^2\alpha
_\parallel^{(2)}}{dt^2})= 0
\label{gunnel}
\end{equation}
and
\begin{equation}
\frac{d^2A^{(2)}}{dt^2}= 0.
\label{coloramo}
\end{equation}
The sum of the leading terms of $\alpha _{\perp, 2}$ to all
orders in
$v$, which is a solution of (\ref{forethought}), is the
second term of (\ref{an1}), i.e.:
\begin{equation}
\alpha_\perp^{(2)}=\sqrt{2t}\rho gv 
K_1(\sqrt {\frac 2t}\rho gv).
\label{maticari}
\end{equation}
To lowest order this reduces to
\begin{equation}
\alpha _\perp^{(2)}\simeq t
\end{equation}
in accordance with (\ref{virajpet}). The solutions of
(\ref{gunnel}) and (\ref{coloramo}) that are compatible with
(\ref{mimseyvane}) and  (\ref{virajpet}) are
\begin{equation}
\alpha _\parallel^{(2)}=t+\frac{\rho ^2g^2v^2}{4}(\log \frac{\rho
^2g^2v^2}{2}+2\gamma +\frac 54),\hspace{1 mm}A^{(2)}(t)=-vt. \label{ernakulam}
\end{equation}

At next order (\ref{ali1}), (\ref{ali2}) and (\ref{ali3})  are:
\begin{eqnarray}&&
\frac{d^2}{dt^2}\frac{d\alpha _\perp
^{(4)}}{dt}+\frac 3t\frac{d}{dt}\frac{d\alpha _\perp
^{(4)}}{dt}-\frac{\rho ^2g^2v^2}{2t^3}\frac{d\alpha _\perp
^{(4)}}{dt}
\nonumber\\&&
=\frac{3}{t}\frac{d\alpha
_\perp^{(2)}}{dt}(\frac{d\alpha
_\perp^{(2)}}{dt}-\frac{d\alpha _\parallel^{(2)}}{dt})
\nonumber\\&&
+\frac{\rho ^2g^2v}{2t^3}(3v\alpha _{\perp}^{(2)}+2A^{(2)})
\frac{d\alpha _\perp ^{(2)}}{dt},
\label{hapagornis}
\end{eqnarray}
\begin{equation}
\frac{1}{t^3}\frac{d}{dt}t^3\frac{d^2\alpha _\parallel
^{(4)}}{dt^2}=-\frac{3}{t}((\frac{d\alpha
_\perp^{(2)}}{dt})^2-(\frac{d\alpha
_\parallel^{(2)}}{dt})^2)
\label{haast}
\end{equation}
and
\begin{equation}
\frac{d^2A^{(4)}}{dt^2}=2(\frac{d\alpha _\perp^{(2)}}{dt})^2v.
\label{henderson}
\end{equation}
Here (\ref{hapagornis}) and (\ref{haast}) have an additional term on the
right hand side from the constraint:
\begin{eqnarray}
-\frac{\rho ^2g^2v^2}{2}\frac{1}{t^2}. \label{alsil3}
\end{eqnarray}

Modifying the constraint with an exponential factor, one obtains
exponential falloff for $t\rightarrow 0$ of (\ref{alsil3}) and consequently
of $\alpha _\perp ^{(4)}$, and $A^{(4)}$ has also exponential falloff by
(\ref{henderson}). The   detailed analysis is very similar
to that carried out in \cite{GHN}, the only difference
being that (\ref{alsil3}) does not cancel the
$O(t^{-2})$ term on the right hand side of
(\ref{hapagornis}). In contrast,
$\alpha _\parallel^{(4)}$ has power law falloff  by the
following term on the right hand side of (\ref{haast}):
\begin{equation}
\frac{3}{t}(\frac{d\alpha _\parallel^{(2)}}{dt})^2
\end{equation}
whence:
\begin{equation}
\alpha _\parallel^{(4)}=\frac 12t^2+\cdots .
\end{equation}
This summation procedure can clearly be iterated

\section{Coupled equations for fermionic zero modes}
\setcounter{equation}{0}
\subsection{General setup}

From (\ref{lagslut}) one obtains the
following coupled equations for the fermionic zero modes:
\begin{equation}
(\sigma\cdot
D )^{ab}\lambda _B^b+g\epsilon ^{abc}\sqrt{2}A^b q^c_A=0 
\label{lambadadadada}
\end{equation}
and
\begin{equation} 
(\bar{\sigma }\cdot D)^{ab} q^b_A-g\epsilon ^{abc}\sqrt{2}A^b\lambda
^c_B=0  
\label{tildegui}
\end{equation}
and a second set of equations obtained by the substitution
$(\lambda _B,q_A)\rightarrow (q_B,-\lambda _A)$.

\subsection{The supersymmetric zero mode}

For the supersymmetric zero mode the
following Ansatz for the gluino field $\lambda _B$ is used:
\begin{equation}
\lambda _B^a=f_\perp(t)\bar{\sigma }\cdot x\sigma
^a\sigma \cdot x u_\sigma, \hspace{1 mm}a=1,2 \label{SOS}
\end{equation}
and
\begin{equation}
\lambda _B^3=f_\parallel(t)\bar{\sigma }\cdot x\sigma
^3\sigma \cdot x u_\sigma, \hspace{1 mm}a=1,2, \label{S3S}
\end{equation}
where $u_\sigma $ is a constant unity twospinor.
The quark field $q_A$ only has transverse components with
the Ansatz:
\begin{equation}
q_A^a=\phi (t)\epsilon ^{ab}\sigma ^b\sigma \cdot xu_\sigma.
\end{equation}
From (\ref{lambadadadada}) and (\ref{tildegui}) the
following set of coupled equations is then obtained:
\begin{eqnarray}&&
6f_\perp(t)-2t\frac{df_\perp(t)}{dt}
-2t\frac{d\log \alpha
_\parallel (t)}{dt}f_\perp(t)
\nonumber\\&&-2t\frac{d\log \alpha
_\perp (t)}{dt}f_\parallel(t)+g\sqrt {2}A(t)\phi(t)=0,
\label{rip}
\end{eqnarray}
\begin{equation}
6f_\parallel(t)-2t\frac{df_\parallel(t)}{dt}
-4t\frac{d\log \alpha
_\perp (t)}{dt}f_\perp(t)=0
\label{rap}
\end{equation}
and
\begin{equation}
-\frac{2t^2}{\rho ^2}(\frac{d\phi (t)}{dt}
-\frac{d\log \alpha
_\parallel (t)}{dt}\phi(t))+g\sqrt {2}A(t)f_\perp(t)=0.
\label{rup}
\end{equation}
These equations are
solved by iteration in the parameter $v$, with even orders
for 
$f_\perp$ and $f_\parallel$ and odd orders for  $\phi $, and
with the order of $v$ indicated by a subscript.

At zeroth order (\ref{rip}) and (\ref{rap}) reduce
to:
\begin{eqnarray}&&
6f_{\perp,0}(t)-2t\frac{df_{\perp,0}(t)}{dt}
-\frac{2t}{1+t}(f_{\perp,0}(t)+f_{\parallel,0}(t))
\nonumber\\&& =6
f_{\parallel,0}(t)-2t\frac{df_{\parallel,0}(t)}{dt}
-\frac{4t}{1+t}f_{\perp,0}(t)=0
\end{eqnarray}
with the solution
\begin{equation}
f_{\perp, 0}(t)=f_{\parallel, 0}(t)
=\frac{t^3}{(1+t)^2}.
\label{ursus}
\end{equation}
At first order  (\ref{rup}) implies:
\begin{equation}
-\frac{2t^2}{\rho^2}(\frac{d\phi_1(t)}{dt}-\frac{\phi
_1(t) }{1+t})=-g\sqrt 2v(\frac{t}{1+t})^3.
\end{equation}
By integration follows:
\begin{equation}
\phi _1(t)=g\sqrt 2v\rho^2\frac 14\frac{t^2+\frac
13t^3}{(1+t)^2} \label{ursula}
\end{equation}
where an integration constant was chosen to make $\phi (t)$
vanish for $t\rightarrow 0$.

At second order the following equations arise from
(\ref{rip}) and (\ref{rap}):
\begin{eqnarray}&&
6f_{\perp,2}(t)-2t\frac{df_{\perp,2}(t)}{dt}
-\frac{2t}{1+t}f_{\perp,2}(t)
-\frac{2t}{1+t}f_{\parallel,2}(t)
\nonumber\\&&
=\frac{2t^4}{(1+t)^2}(\frac{d}{dt}\frac{\alpha
_{\perp,2}}{1+t}+\frac{d}{dt}\frac{\alpha
_{\parallel ,2}}{1+t})-\frac{\rho
^2g^2v^2}{2}\frac{t^2+\frac 13t^3}{(1+t)^3}
\label{talpra}
\end{eqnarray}
and
\begin{equation}
6f_{\parallel,2}(t)-2t\frac{df_{\parallel,2}(t)}{dt}
-\frac{4t}{1+t}f_{\perp,2}(t)
=\frac{4t^4}{(1+t)^2}\frac{d}{dt}\frac{\alpha_{\perp,2}}{1+t}
\label{magyar}
\end{equation}
whence
\begin{eqnarray}&&
\frac{d}{dt}
\frac{(1+t)^2}{t^3}(2f_{\perp,2}(t)+f_{\parallel,2}(t)) 
\nonumber\\&&
=\frac13\frac{\rho
^2g^2v^2}{2t^2}(1+\frac{2}{1+t})-4\frac{d}{dt}\frac{2\alpha
_{\perp, 2}+\alpha _{\parallel,
2}}{1+t}
\label{bosnaseraj} 
\end{eqnarray}
and
\begin{equation}
\frac{d}{dt}\frac{f_{\perp ,2}(t)-f_{\parallel
,2}(t)}{t^3(1+t)} =\frac{\Psi _2(t)}{(1+t)^2 }+\frac{\rho
^2g^2v^2}{4}\frac{1+\frac 13t}{t^2(1+t)^4}
\label{marica}
\end{equation} 
that are both solved by quadrature.
For $t\rightarrow \infty$ the right hand sides of
(\ref{bosnaseraj}) and (\ref{marica}) are $O(t^{-2})$ and
$O(t^{-5})$, respectively. Thus it follows by integration of
these equations  that both $f_{\perp,2} (t)$ and
$f_{\parallel,2}(t)$ are $O(t^{0})$ in this limit. In the
limit
$t\rightarrow 0$ it follows from the two equations that
both 
$f_{\perp,2} (t)$ and
$f_{\parallel,2}(t)$ are $O(t^2)$. The supersymmetric zero
mode is therefore square integrable also in second order.

\subsection{Asymptotic behaviour}

A  proof by induction similar to that given in the bosonic
case is carried out on the basis  of (\ref{rip}),
(\ref{rap}) and (\ref{rup}) that 
$\frac{\phi (t)}{ t}$, $f_\perp(t)$ and
$f_\parallel(t)$  at most have logarithmic growth for
$t\rightarrow \infty $ at each order in $v$, except
$f_{\perp,0}$ and  $f_{\parallel,0}$ which grow
linearly.  Assuming the above-mentioned estimates hold
to orders less than $n$, one sees immediately from
(\ref{rip}) and (\ref{rap}) that $f_{\perp,0}$ and 
$f_{\parallel,0}$ are
$O(t^0)$ for
$t\rightarrow \infty$. From  (\ref{rup}) we get:
\begin{equation}
\frac{d\phi_n(t)}{dt}-\frac{\phi _n(t)}{1+t}
\propto t^{-2}
\end{equation} 
establishing the estimate for $\phi $ also to order $n$.
Thus the estimate holds to all orders.

Then  the limit $t\rightarrow 0$ is investigated.
For the leading terms for $t\rightarrow 0$ one has the estimate
\begin{eqnarray}
f_{\perp, n} \propto
t^{3-\frac{n}{2}},\hspace{2mm}
  \phi_{ n}(t)\propto t^{\frac{5}{2}-\frac{n}{2}}. 
\label{sparkst¿tting}
\end{eqnarray}
The analysis of the leading, nextleading etc. terms of $f$
and
$\phi $  is quite similar to that carried  out on the
corresponding quantities for supersymmetric QCD in
\cite{GHN}, to which we refer for details. The new feature
is the presence of $f_\parallel (t)$, which by (\ref{rap})
and (\ref{ursus}) has the leading term in the double series
expansion of $\rho $ and $\sqrt t$:
\begin{equation}
f_\parallel^{(6)}(t)=t^3
\label{onyxsenator}
\end{equation}
while the nextleading terms are determined by
\begin{equation}
-2t^4\frac {d}{dt}\frac{f_\parallel ^{(8)}(t)}{t^3}
=4t\frac{d\alpha _\perp ^{(2)}(t)}{dt}f_\perp^{(6)}(t)
\label{ecclesiastes}
\end{equation}
with the solution
\begin{equation}
f_\parallel^{(8)}(t)=-2t^3\int
_0^t\frac{dt'}{(t')^3}\frac{d\alpha _\perp
^{(2)}(t')}{dt}f_\perp^{(6)}(t')
\label{archimandrit}
\end{equation}
which in lowest order of the expansion in $v$ agrees with
the $O(t^4)$ term of (\ref{ursus}). The right-hand side of
(\ref{ecclesiastes}) shows exponential decrease for
$t\rightarrow 0$ ($x\rightarrow \infty $) after summation over
all orders of $v$, and so does therefore $f^{(8)}(t)$ by
(\ref{archimandrit}), in contrast to the leading term that
by (\ref{onyxsenator}) has a power law decrease for
$x\rightarrow \infty$. A similar phenomenon was observed
for the scalar field $A$.

\subsection{The superconformal zero mode}

To obtain the superconformal zero mode the equations
(\ref{lambadadadada}) and (\ref{tildegui}) are solved by the  following Ansatz
for the gluino field:
\begin{equation}
\lambda _B^a=f_\perp(t)\bar{\sigma }\cdot x\sigma
^a u_\sigma, \hspace{1 mm}a=1,2 \label{SCS}
\end{equation}
and
\begin{equation}
\lambda _B^3=f_\parallel(t)\bar{\sigma }\cdot x\sigma
^3 u_\sigma, \hspace{1 mm}a=1,2 \label{SPS}
\end{equation}
and for the quark field:
\begin{equation}
q_A^a=\phi (t)\epsilon ^{ab}\sigma ^bu_\sigma; \hspace{1 mm}a,b=1,2
\end{equation}
where the functions $f_\perp(t)$, $f_\parallel (t)$ and $\phi (t)$ obey the
coupled equations:
\begin{eqnarray}&&
4f_\perp(t)-2t\frac{df_\perp(t)}{dt}
-2t\frac{d\log \alpha
_\parallel (t)}{dt}f_\perp(t)
\nonumber\\&&-2t\frac{d\log \alpha
_\perp (t)}{dt}f_\parallel(t)+g\sqrt {2}A(t)\phi(t)=0,
\label{reep}
\end{eqnarray}
\begin{equation}
4f_\parallel(t)-2t\frac{df_\parallel(t)}{dt}
-4t\frac{d\log \alpha
_\perp (t)}{dt}f_\perp(t)=0
\label{raap}
\end{equation}
and
\begin{equation}
-\frac{2t^2}{\rho ^2}(\frac{d\phi (t)}{dt}
-\frac{d\log \alpha
_\parallel (t)}{dt}\phi(t))+g\sqrt {2}A(t)f_\perp(t)=0.
\label{roop}
\end{equation}

At zeroth order the solution of (\ref{reep}) and (\ref{raap}) is:
\begin{equation}
f_{\perp, 0}(t)=f_{\parallel, 0}(t)=\frac{t^2}{(1+t)^2}.
\end{equation}
At first order (\ref{roop}) is:
\begin{equation}
-\frac{2t^2}{\rho ^2}(\frac{d\phi _1(t)}{dt}-\frac{\phi
_1(t)}{1+t})+g\sqrt 2v\frac{t^2}{(1+t)^3}=0
\end{equation}
with the solution
\begin{equation}
\phi _1(t)=-\frac 16\frac{g\sqrt 2v\rho
^2}{(1+t)^2}
\label{asterix}
\end{equation}
where an integration constant has to vanish for
the sake of the asymptotic behaviour for $t\rightarrow
\infty $.

Adding to $\phi _1(t)$ a term 
\begin{equation}
\phi _{1, {\rm add}}=\frac 16g\sqrt 2v\rho^2
\end{equation}
in order to obtain a solution that vanishes for $t\rightarrow 0$, means
having a nonzero right-hand side of (\ref{tildegui}):
\begin{equation}
 \frac{t^2}{1+t}\epsilon ^{ab}\bar{\sigma }\cdot x\sigma
^bu_\sigma \simeq \frac{1}{3}g\sqrt 2v(1+t)\epsilon
^{ab}\lambda ^b_B
\label{schweber}
\end{equation} 
where the last version of (\ref{schweber}) is produced by an additional Yukawa
coupling term in (\ref{lagslut}):
\begin{equation}
\frac{1}{3}g\sqrt 2v(1+t)(q^a_B)^\dagger \epsilon ^{ab}
\lambda ^b_B.
\end{equation}

For $t\rightarrow 0$ the analysis of leading, nextleading
etc. terms is very similar to the corresponding analysis
of supersymmetric QCD carried out in \cite{GHN}.
In the other limit, $t\rightarrow \infty$, it follows from
(\ref{reep}), (\ref{raap}) and (\ref{roop}) that 
$f_{\perp,0}\rightarrow 1,
f_{\parallel,0}\rightarrow 1$ while 
$f_{\perp,n}\propto t^{-1},
f_{\parallel,0}\propto t^{-1}, n\neq 0$ and $\phi
\propto t^{-2}$.

\section{Conclusion}

The main resultat of this paper is in a sense a negative one:
Despite the considerable complications arising from the
presence of two different gauge prepotentials, while
\cite{NN}, \cite{GHN} had only one, the result of the
analysis is exactly the same: Constraints are required
that produce additional terms in the gauge field equations
at second and fourth order of the gauge breaking parameter
$v$. 

It is perhaps less surprising that also  the outcome of the
analysis of the
fermionic zero modes is the same as in \cite{GHN}: While
the supersymmetric zero mode is perfectly well behaved,
the superconformal zero mode has a nonpermissible large
distance behaviour at first order in $v$ that is eliminated
by an additional Yukawa coupling. 

\section{ Acknowledgement} A helpful discussion with Professor
U. Haagerup is gratefully acknowledged.
\appendix
\setcounter{equation}{0}

\section{Solution of the inhomogeneous hypergeometric equation}

The starting point is the following two differential equations
fulfilled  according to (\ref{ali1})-(\ref{ali2}) by the gauge field
prepotentials
$\alpha _{\perp}$ and
$\alpha _{\parallel}$ in a general order of the
expansion in the parameter $v$:  
\begin{eqnarray}&&
(1+t)^2\frac{d}{dt}\left((\frac{t}{1+t})^3
\frac{d^2\alpha _{\perp}}{dt^2}\right)
\nonumber\\&&-
\frac{t^2}{1+t}(3-\frac{2t}{1+t})\frac{d}{dt}\frac{\alpha
_{-}}{1+t}=\chi _{\perp}
\label{aliutor}
\end{eqnarray}
and
\begin{eqnarray}&&
(1+t)^2\frac{d}{dt}\left((\frac{t}{1+t})^3
\frac{d^2\alpha _{\parallel}}{dt^2}\right)
\nonumber\\&&
+\frac{t^2}{1+t}(6-\frac{4t}{1+t})\frac{d}{dt}\frac{\alpha
_{-}}{1+t}=\chi _{\parallel}
\label{alianora}
\end{eqnarray}
with $\alpha _-=\alpha _\perp-\alpha _\parallel $,  and the function
$\Psi $ is defined as:
\begin{equation}
\Psi =\frac{1}{1+t}\frac{d}{dt}\frac{\alpha _-}{1+t}.
\end{equation}
By combination of (\ref{aliutor})  and (\ref{alianora})
follows:
\begin{eqnarray}&&
(1+t)t^3\frac{d^2\Psi}{dt^2}+(3t^2(1+t)+2t^3)\frac{d\Psi}{dt}
\nonumber\\&&
=\frac{1}{1+t}\frac{d}{dt}t^3(1+t)^2\frac{d\Psi}{dt}  
=\chi_\perp-\chi_\parallel.
\label{cypresmord}
\end{eqnarray}

This equation can obviously be solved by quadrature,
but it is convenient to rewrite it as an
inhomogeneous hypergeometric equation. Switching to the
variable
$u=\frac{t}{1+t}$  one finds:
\begin{equation}
u(1-u)    \frac{d^2\Psi }{du^2}+3\frac{d\Psi
}{du} =X
\label{tutankhamon}
\end{equation}
with
\begin{equation}
X=\frac{1+t}{t^2}(\chi _\perp-\chi_\parallel).
\end{equation}

The
homogeneous equation:
\begin{equation}
u(1-u)\frac{d^2\Psi}{du^2}+3\frac{d\Psi}{du}=0
\label{astaire}
\end{equation}
has the independent solutions 
\begin{equation}
F(u)=1
\label{genekelly}
\end{equation}
and
\begin{equation}
G(u)=-\frac{1}{2u^2}+\frac{3}{u}+3\log u-u-\frac 32
\label{azzuwaza}
\end{equation}
with
\begin{equation}
\frac{dG(u)}{du}=(\frac{1-u}{u})^3;\hspace{1 mm}G(1)=0.
\label{johannkepler}
\end{equation}

A particular solution of
the  inhomogeneous equation 
(\ref{tutankhamon})  is, with
$0<u_0\leq 1$:
\begin{eqnarray}&&
\Psi (u)
=\int
_{u_0}^u(G(u)-G(u'))(u')^2
(1-u')^{-4}X(u')du'
\nonumber\\&&
\label{woowoowoo}
\end{eqnarray}
to which is added a linear combination of $F(u)$ and
$G(u)$ in order to obtain the most general solution. $X(u)$ is $O((1-u)^3)$ near
$u=1$ for all orders of $v$  
and the first integral in (\ref{woowoowoo})  thus
diverges at most logarithmically  for 
$u\rightarrow 1$. The integration constants are chosen such
that the final solution of (\ref{tutankhamon}) is
\begin{eqnarray}&&
\Psi(u)=\int
_{u}^1G(u')(u')^2(1-u')^{-4}X(u')du'
\nonumber\\&&
+(\int
_{u_0}^u(u')^2(1-u')^{-4}X(u')du'+C_G)G(u).
\nonumber\\&&
\label{weeweewee}
\end{eqnarray}
with $C_G$ an integration constant, and it follows from
(\ref{johannkepler}) that
$\Psi(u)$ is $O((1-u)^4)$ near $u=1$.

In  general the leading term at $n$th order for
$t\rightarrow 0 \hspace{1 mm} (u\rightarrow 0)$ 
is: 
\begin{equation}
X _{n}\propto t^{-1-\frac n2}.
\label{sultan}
\end{equation}
From this estimate  then follows  
near $u=0$:
\begin{equation}
\Psi _n(u)\propto u^{-\frac n2}.
\label{zolikon}
\end{equation}

\end{document}